\documentclass[twoside,fleqn]{article}

\usepackage{epsf,espcrc2}
\usepackage[hyperindex]{hyperref}
\newcommand\eprint[1]{\publishedeprint{#1}{#1}}
\newcommand\publishedeprint[2]{\href{http://xxx.lanl.gov/abs/#2}{#1}}
\ifx\href\undefined\newcommand\href[2]{#1}\fi

\newcommand{\beq}{\begin{equation}}
\newcommand{\eeq}{\end{equation}}
\newcommand{\beqa}{\begin{eqnarray}}
\newcommand{\eeqa}{\end{eqnarray}}

\newcommand\0{\hphantom{0}}

\newskip\defaultbaselineskip\defaultbaselineskip=12pt

\def\GeV{\mathord{\rm \;GeV}}
\def\MeV{\mathord{\rm \;MeV}}

\def\bar{\overline}

\def\gsim{\mathrel{\raise2pt\hbox to 8pt{\raise -5pt\hbox{$\sim$}\hss{$>$}}}}
\def\rsim{\mathrel{\raise2pt\hbox to 8pt{\raise -5pt\hbox{$\sim$}\hss{$>$}}}}
\def\lsim{\mathrel{\raise2pt\hbox to 8pt{\raise -5pt\hbox{$\sim$}\hss{$<$}}}}
\def\ssqr#1#2{{\vbox{\hrule height.#2pt
      \hbox{\vrule width.#2pt height#1pt \kern#1pt\vrule width.#2pt}
      \hrule height.#2pt}\kern-.#2pt}}

\def\NPBPS#1{{\it Nucl. Phys.} {\bf B} ({\it Proc. Suppl.}) {\bf #1}}
\def\PRL#1{{\it Phys. Rev. Lett.} {\bf #1}}

\def\PLB#1{{\it Phys. Lett.} {\bf #1B}}

\def\PREP#1{{\it Phys. Rep.} {\bf #1}}

\def\etal{{\it et al.\ }}

\def\gsim{\mathrel{\raise2pt\hbox to 8pt{\raise -5pt\hbox{$\sim$}\hss{$>$}}}}
\def\rsim{\mathrel{\raise2pt\hbox to 8pt{\raise -5pt\hbox{$\sim$}\hss{$>$}}}}
\def\lsim{\mathrel{\raise2pt\hbox to 8pt{\raise -5pt\hbox{$\sim$}\hss{$<$}}}}

\newcommand\figcaption[1]{\vskip-0.0truein\caption{#1}\vskip0.0truein}
\newcommand\alphamsbar{\hbox{$\alpha_{\overline{MS}}$}}

\newcommand\MSbar{\hbox{$\overline{MS}$}}
\newcommand\mmsbar{\hbox{$m_{\overline{MS}}$}}
\newcommand\mbar{\hbox{$\overline{m}$}}
\def\cpt{\hbox{$ \chi $PT}}

\title{Light Quark Masses and the CP violation parameter $\epsilon'/\epsilon$}

\author{R. Gupta and T.~Bhattacharya\address{T-8 Group, MS B285, 
Los Alamos National Laboratory, Los Alamos, New Mexico 87545 U.S.A.}
\thanks{We acknowledge the support of the ACL at Los Alamos and 
NCSA at Urbana for computational support.}}

\begin{document}

\begin{abstract}
We present estimates of light quarks masses using lattice data.  Our
main results are based on a global analysis of all the published data
for Wilson and Staggered fermions, both in the quenched approximation
and with $n_f=2$ dynamical flavors.  The Wilson and Staggered results
agree after extrapolation to the continuum limit for both the
$n_f=0,2$ theories.  Our best estimates, in the \MSbar\ scheme at
scale $2 \GeV$, are $\mbar=3.2(4) \MeV$ and $m_s = 90(20)
\MeV$ in the quenched approximation, and $\mbar \sim 2.7 \MeV$ and
$m_s \sim 70 \MeV$ for the $n_f=2$ theory.  These estimates are
significantly smaller than phenomenological estimates based on sum
rules, but maintain the ratios predicted by chiral perturbation theory
(\cpt). Along with the new estimates of 4-fermion operators, lower
quark masses have a significant impact on the extraction of
$\epsilon'/\epsilon$ from the Standard Model.
\end{abstract}

\maketitle

\section{LIGHT QUARK MASSES}
\label{s_mq}

The masses of light quarks $m_u$, $m_d$, and $m_s$ are three of the
least well known parameters of the Standard Model. These quark
masses have to be inferred from the masses of low lying hadrons.
\cpt\ relates the masses of pseudoscalar mesons to $m_u,\ m_d$, and 
$m_s$, however, the presence of the unknown scale $\mu$ in ${\cal
L}_{\cpt}$ implies that only ratios of quark masses can be determined.
For example $2m_s/(m_u + m_d) \equiv m_s/\mbar= 25$ at lowest order,
and $31$ at next order \cite{gasserPR,Donoghue}.  Latest estimates using QCD
sum rules give $m_u + m_d = 12(1) \MeV$ \cite{BPR95}.  However, as
discussed in \cite{rSUMRULE96BGM}, a reanalysis of sum rules shows
that far more experimental information on the hadronic spectral
function is needed before sum rules can give reliable estimates.
Thus, lattice QCD is currently the most promising approach.

To extract $a$, $\mbar$, $m_s$ we fit the global data as 
\begin{eqnarray}
M_{PS} &=&           B_{PS}  (m_1 + m_2)/2  \nonumber \\
M_{V}  &=& A_{V}   + B_{V}   (m_1 + m_2)/2  \ ,
\label{e:chiralfits}
\end{eqnarray}
for each value of the lattice parameters, $\beta$, $n_f$, fermion
action. From $B_{PS}, A_V, B_V$ we determine the three desired
quantities; the lattice scale $a$ using $M_\rho$, $\mbar$ using
$M_\pi^2 / M_\rho^2$, and $m_s$ in three different ways using $M_K, \
M_{K^*}, \ M_\phi$. Throughout the analysis we assume that $\phi$ is a
pure $s \bar s$ state.  Note that using 
Eq.~\ref{e:chiralfits} means that we can predict
only one independent quark mass from the pseudoscalar data, which we
choose to be $\mbar$. The reason for this truncation is that in most
cases the data for $M_\pi$ and $M_\rho$ exist at only $2-4$ values of
``light'' quark masses in the range $0.3m_s - 2m_s$.  In this
restricted range of quark masses the existing data do not show any
significant deviation from linearity. One thus has to use $M_V$ in 
order to extract $m_s$. Details of our analysis 
and of the global data used are given in \cite{rMq96LANL}. 

For Wilson fermions the lattice quark mass, defined at scale $q^*$, is
taken to be $m_L(q^*) a = ({1 / 2\kappa} - {1/ 2 \kappa_c} )$.  For
staggered fermions $m_L(q^*) = m_0$, the input mass. The \MSbar\ mass
at scale $\mu$ is $\mmsbar(\mu) = Z_m(\mu a) m_L(a)$, where $Z_m$ is
the mass renormalization constant relating the lattice and the
continuum regularization schemes at scale $\mu$, and $\lambda =
g^2/16\pi^2$.  In calculating $Z_m$, $a\ la$ Lepage-Mackenzie, we use
$\alphamsbar$ for the lattice coupling, use ``horizontal'' matching,
$i.e.$ $\mu = q^*= 1/a$, and do tadpole subtraction. We find that the
results are insensitive to the choice of $q^*$ in the range $0.86/a -
\pi/a$ and to whether or not tadpole subtraction is done.  Once
$\mmsbar(\mu)$\ has been calculated, its value at any other scale $Q$
is given by the two loop running. We quote all results at $Q=2 \GeV$.

We extrapolate the lattice masses to $a=0$ using the lowest order
corrections (Wilson are $O(a)$ and Staggered are $O(a^2)$).  In the quenched 
fits we omit points at the stronger couplings ($a > 0.5 \GeV^{-1}$)
because we use only the leading correction in the extrapolation to
$a=0$, and because the perturbative matching becomes less reliable as
$\beta$ is decreased.  The bottom line is that we find that the leading
corrections give a good fit to the data, and in the $a =0$ limit the
two different fermion formulations give consistent results.  Our final 
results are summarized in Table \ref{t_m}. 

\begin{table} 
\caption{Summary of results in $\MeV$ in $\MSbar$ scheme at $\mu=2\ \GeV$. 
The label $W(0)$ stands for Wilson with $n_f=0$. An additional
uncertainty of $\sim 10\%$ due to the uncertainty in the lattice scale
$a$ is suppressed.\looseness=-1}
\def\q{\quad}
\def\s{\phantom{-}}
\def\sq{\phantom{-}\q}
\setlength{\tabcolsep}{3pt}
$$
\begin{tabular}{|l|c|c|c|c|}
\hline
$         $&$ \mbar    $&$ m_s(M_K)  $&$ m_s(M_\phi)  $&$ m_s(M_K^*)  $\cr
\hline		
$ W(0)    $&$ 3.3(4)   $&$ 83(10)    $&$ 96(10)       $&$ 76(20)      $\cr
$ S(0)    $&$ 3.1(1)   $&$ 78(\0 3)  $&$ 96(\0 2)     $&$ 87(\0 2)    $\cr
\hline
$ W(2)    $&$ 2.5(3)   $&$ 63(8)     $&$ 77(10)       $&$ 78(22)      $\cr
$ S(2)    $&$ 2.9(3)   $&$ 73(8)     $&$ 66(\0 6)     $&$ 59(\0 6)    $\cr
\hline
\end{tabular}
$$

\label{t_m}
\end{table}

\begin{figure}[t]
\hbox{\hskip15bp\epsfxsize=0.9\hsize \epsfbox {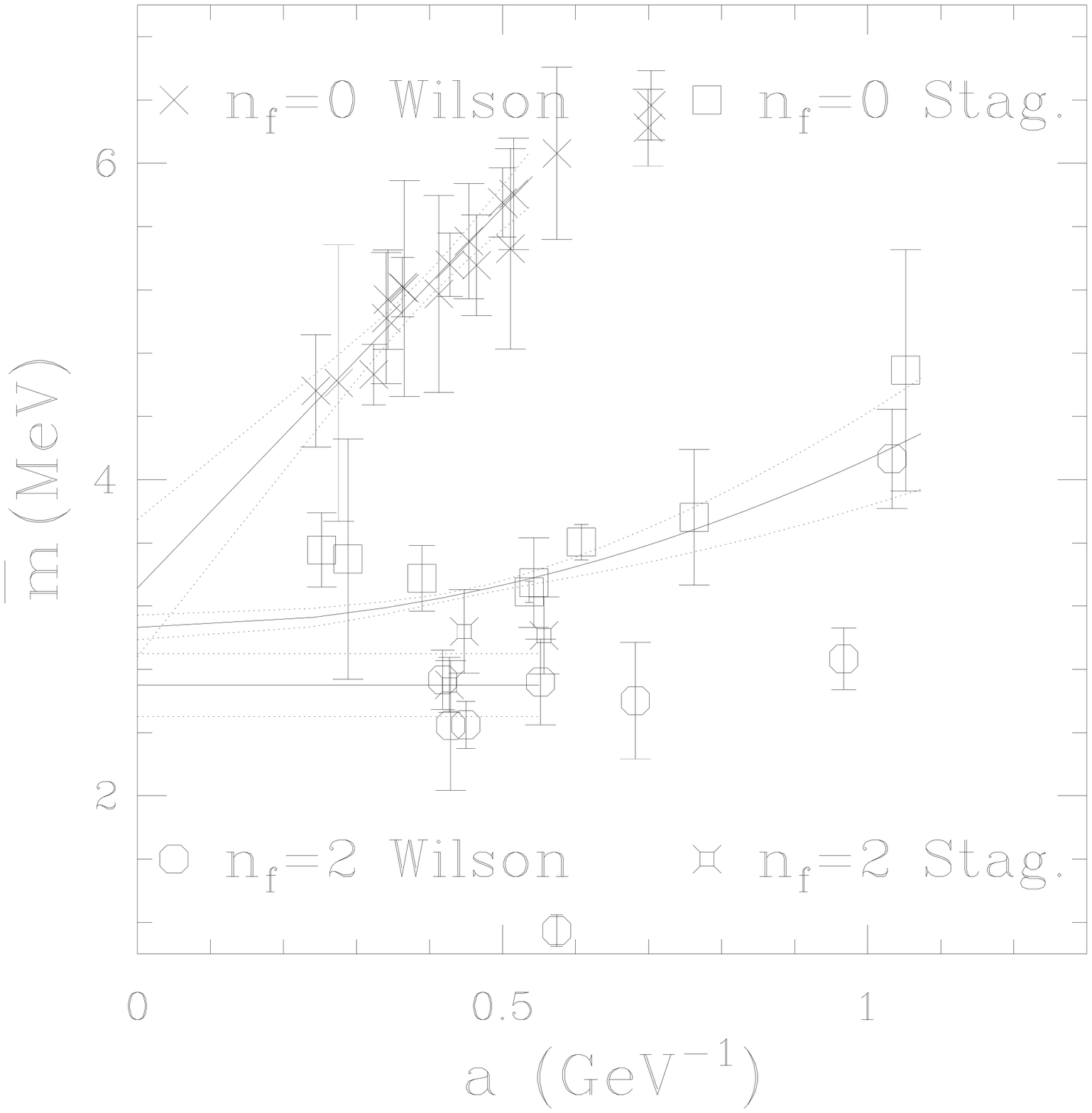}}
\figcaption{$\mbar(\MSbar,2 \GeV)$ extracted using $M_\pi$ data with the scale 
set by $M_\rho$.}
\label{f_mbar}
\end{figure}

The global data for $\mbar$ and the extrapolations to $a=0$ for Wilson
are shown in Fig.~\ref{f_mbar}.  Using the average of quenched estimates 
given in Table~\ref{t_m} we get 
\begin{equation}
\mbar(\MSbar,2 \GeV) = 3.2(4)(3)  \MeV \quad {\rm (quenched)} .
\end{equation}
where the first error estimate is the larger of the two extrapolation 
errors, and the second is that due to the uncertainty in the scale $a$. 

The pattern of $O(a)$ corrections in the unquenched data ($n_f =2$) is
not clear and we only consider data for $\beta \ge 5.4$. The strongest
statement we can make is qualitative; at any given value of the
lattice spacing, the $n_f=2$ data lies below the quenched result.
Taking the existing data at face value, we find that the average of
the Wilson and staggered values are the same for the choices $\beta
\ge 5.4$, $\beta \ge 5.5$, or $\beta \ge 5.6$.  We therefore take this
average
\begin{equation}
\mbar(2 \GeV) \approx  2.7  \MeV \qquad (n_f=2 {\rm\ flavors}) \ ,
\end{equation}
as the current estimate. To obtain a value in the physical case
of $n_f=3$, the best we can do is to assume a behavior linear in $n_f$. 
In which case extrapolating the $n_f=0$ and $2$ data gives
\begin{equation}
\mbar(2 \GeV) \approx 2.5  \MeV \qquad (n_f=3 {\rm\ flavors}) .
\end{equation}
We stress that this extrapolation in $n_f$ is extremely preliminary. 

\begin{figure}[t]
\hbox{\hskip15bp\epsfxsize=0.9\hsize \epsfbox {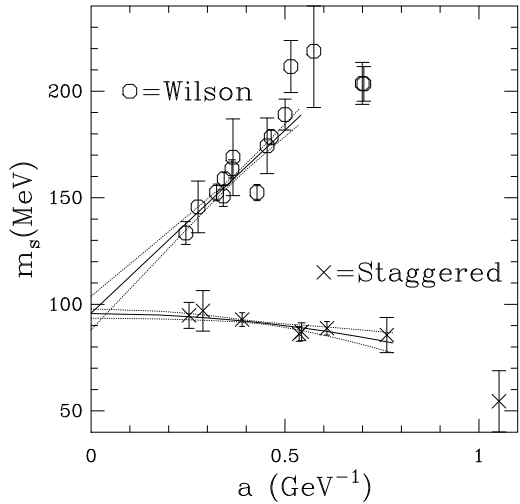}}
\figcaption{Comparison of $m_s(\MSbar,2 \GeV)$ extracted using $M_\phi$ 
for the quenched Wilson and staggered theories.}
\label{f_msphiQ}
\end{figure}

We determine $m_s$ using the three different mass-ratios,
$M_K^2/M_\pi^2$, \ $M_{K^*} / M_\rho $, and $M_{\phi} / M_\rho$.
Using a linear fit to the pseudo-scalar data constrains ${m_s(M_K) =
25 \mbar }$. Using the vector mesons $M_K^*$ and $ M_\phi$ gives
independent estimates.  The quenched data and extrapolation to $a=0$
of $m_s(M_\phi)$ are shown in Fig.~\ref{f_msphiQ}. The average of
Wilson and staggered values are
$m_s(M_\phi)  = 96(10)  \MeV $ and $m_s(M_{K^*}) = 82(20)  \MeV$
where the errors are taken to be the larger of Wilson/staggered data. 
From these we get our final estimate 
\begin{equation}
m_s = 90(15)  \MeV \qquad ({\rm quenched}) \ .
\end{equation}
The $n_f=2$ data shows a pattern similar to that for $\mbar$.
Therefore, we again take the average of values quoted in Table~\ref{t_m} 
to get
\begin{equation}
m_s = 70(11)  \MeV \qquad (n_f=2) \ . 
\end{equation}
The error estimate reflects the spread in the data. 

Qualitatively, the data show three consistent patterns. First, agreement 
between Wilson and Staggered values. Second, for a
given value of $a$ the $n_f=2$ results are smaller than those in the
quenched approximation.  Lastly, the ratio $\mbar / m_s(M_\phi)$ is in
good agreement with the next-to-leading-order predictions of chiral
perturbation theory for both the $n_f=0$ and $2$ estimates.
It is obvious that more lattice data are needed to resolve the
behavior of the unquenched results. However, the surprise of this
analysis is that both the quenched and $n_f=2$ values are small and
lie at the very bottom of the range predicted by phenomenological
analyses \cite{gasserPR}.

\section{CP VIOLATION and $\epsilon'/\epsilon$}
\label{s_CP}

A detailed analysis of 4-fermion matrix elements with quenched Wilson
fermions at $\beta=6.0$ is presented in \cite{rBK96LANL}. The 
methodology, based on the expansion of the matrix elements in powers of 
the quark mass and momentum, is discussed in \cite{rBK95LAT}. Our 
estimates in the NDR scheme at $\mu=2\ \GeV$ are 
\begin{eqnarray}
B_K       &=& 0.68(4) \ , \nonumber \\
B_D       &=& 0.78(1) \ , \nonumber \\
B_7^{3/2} &=& 0.58(2) \ , \nonumber \\
B_8^{3/2} &=& 0.81(3) \ .
\end{eqnarray}
The errors quoted are statistical. The major remaining sources of 
errors in these  estimates are lattice discretization and quenching. 

To exhibit the dependence of the Standard Model (SM) prediction 
of $\epsilon'/\epsilon$ on the light quark masses and the $B$ parameters
we write 
\begin{equation} 
\epsilon'/\epsilon = A \bigg(c_0 + c_6 B_6^{1/2} M_r + c_8 B_8^{3/2} M_r \bigg) \ ,
\end{equation}
where $M_r = (158\MeV/(m_s + m_d))^2$. For reasonable choices of SM
parameters Buras \etal\ estimate
$A = 1.3\times 10^{-4}$, $c_0 = - 1.3$, $c_6 = 7.9$, $c_8 = - 4.3$
\cite{rCP96Buras}. Thus, to a good approximation 
$\epsilon'/\epsilon \propto M_r$; and increases as $B_8^{3/2}$
decreases.  As a result, our estimates of $\mbar, m_s, B_8^{3/2}$
increase $\epsilon'/\epsilon $ by roughly a factor of three compared
to previous analysis, $i.e.$ from $3.6\times10^{-4}$ to $\sim
10.4\times10^{-4}$. This revised estimate lies in between the Fermilab
E731 ($7.4(5.9)\times10^{-4}$) and CERN NA31 ($23(7)\times10^{-4}$)
measurements and provides a scenario in which direct CP violation can
be explained within the Standard Model.


\begin{thebibliography}{19}
%
\bibitem{gasserPR} J.~Gasser, H.~Leutwyler, \PREP{C87} (1982) 77.
\bibitem{Donoghue} J. Donoghue, B. Holstein, D. Wyler, \PRL{69} (1992) 3444.
%
%
\bibitem{BPR95}
        J.~Bijnens, J.~Prades, and E.~de Rafael, \PLB{348} (1995) 226. 

\bibitem{rSUMRULE96BGM}
        T.~Bhattacharya, R.~Gupta, K.~Maltman, Los Alamos Preprint LA-UR-96-2698. 
\bibitem{rMq96LANL}
        T.~Bhattacharya, R.~Gupta, \eprint{hep-lat/9605039}. 

\bibitem{rBK96LANL}
        R.~Gupta, T.~Bhattacharya, S.~Sharpe, LAUR-96-1829. 
\bibitem{rBK95LAT}
        R.~Gupta, T.~Bhattacharya, \NPBPS{47} (1996) 473.
\bibitem{rCP96Buras} A.~Buras, \etal, \eprint{hep-ph/9608365}.
\end{thebibliography}
\end{document}